\documentclass[aps,10pt,twocolumn,showpacs]{revtex4}
\usepackage{amsfonts}
\usepackage{amsmath}
\usepackage{amssymb}
\usepackage{verbatim}
\usepackage[dvips]{graphicx}

\begin{document}

\title{Implementation of a double-scanning technique \\
for studies of the Hanle effect in Rubidium vapor}
\author{A.Atvars$^{1}$, M. Auzinsh$^{1}$, E.A. Gazazyan$^{2}$, A.V. Papoyan$%
^{2}$, S.V. Shmavonyan$^{2}$} \affiliation{$^{1}$Department of
Physics and Institute of Atomic Physics and Spectroscopy, University
of Latvia, 19 Rainis Blvd.,
LV-1586 Riga, Latvia \\
 $^{2}$Institute for Physical Research, NAS of
Armenia, Ashtarak-2, 0203 Armenia  }

\begin{abstract}
We have studied the resonance fluorescence of a room-temperature
rubidium vapor exited to the atomic $^{5}$P$_{3/2}$ state (D$_{2}$
line) by powerful single-frequency cw laser radiation (1.25
W/cm$^{2}$) in the presence of a magnetic field. In these studies,
the slow, linear scanning of the laser frequency across the
hyperfine transitions of the D$_{2}$ line is combined with a fast
linear scanning of the applied magnetic field, which allows us to
record frequency-dependent Hanle resonances from all the groups of
hyperfine transitions including V- and $\Lambda$ - type systems.
Rate equations were used to simulate fluorescence signals for
$^{85}$Rb due to circularly polarized exciting laser radiation with
different mean frequency values and laser intensity values. The
simulation show a dependance of the fluorescence on the magnetic
field. The Doppler effect was taken into account by averaging the
calculated signals over different velocity groups. Theoretical
calculations give a width of the signal peak in good agreement with
experiment.
\end{abstract}

\date{\today }
 \maketitle

\section{Introduction}

Sustained interest in resonant magneto-optical effects in atomic
vapors is due to their importance to fundamental
physics but also  to possibility of using these effects in numerous applications (see \ \cite%
{Mor91,Ale93,Bud02,Ale05} and references therein). Besides the
obvious dependence on the applied magnetic field, the resonant
nature of these effects implies a substantial dependence on the
laser frequency. Meanwhile, as a rule, only one or two of these
parameters is varied in a given experimental measurement. In the
present work, we report the results of an investigation of the
nonlinear Hanle effect \cite{Pap02,Aln03} where both the laser
frequency and the magnetic field were scanned continuously and
simultaneously at different sequence rates \cite{And07}. This
technique enables us to acquire additional information in one
measurement sequence , and can help to better understand and model
the processes under study. This technique can be applied to other
magneto-optical effects as well \cite{Pap06}.

Usually, models of nonlinear magneto-optical effects are based on
the optical
Bloch equations \cite{Ste05}. At the same time, as was shown in \cite%
{Blu04}, simpler rate equations for Zeeman coherences for stationary
or quasi-stationary excitation are equivalent to the optical Bloch
equations. The model based on the rate equations was successfully
applied to studies of atomic interactions with laser radiation in
cells in the presence of an external magnetic field
\cite{Aln01,Pap02,Aln03}.

The elaboration of a realistic model of the interaction of alkali
atoms with laser radiation is complicated by the fact that the fine
structure levels of a typical alkali atom consists of several
hyperfine structure (HFS) levels that in experiments in cells  can
be only partially resolved spectroscopically due to Doppler
broadening. If an external magnetic field is applied, the  situation
becomes even more complicated. The magnetic field mixes together
magnetic sublevels with the same magnetic quantum number $M$, but
belonging to different hyperfine states. This mixing can be quite
strong, as was shown experimentally and confirmed in a model for Rb
atoms \cite{Sar05} confined in an extremely thin cell \cite{Sar01}.
In that study the extremely thin cell allowed to resolve
spectroscopically transitions between specific magnetic sublevels.
When scanning the laser frequency there appeared in a fluorescence
excitation spectrum transitions that would not be allowed in the
absence of the level mixing due to magnetic field.

In the present study we propose to show that in the model of the
nonlinear Hanle effect we have for the first time accounted
\emph{simultaneously} for all these effects, namely, the creation of
Zeeman coherences between magnetic sublevels in the ground and
excited states, the mixing of the different hyperfine levels in the
magnetic field (partial decoupling of the electronic angular
momentum of the electrons from the nuclear spin), and the Doppler
effect in the manifold of magnetic sublevels of all hyperfine levels
of the  fine structure states involved.

\section{Experiment}

\begin{figure}[tbp]
\includegraphics[width=7.5cm]{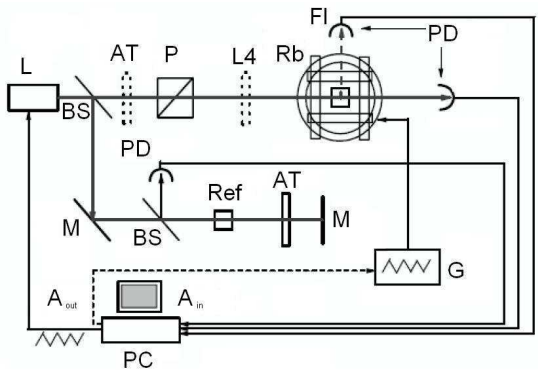}
\caption{Experimental Setup. Abbreviations: L- laser, BS- beam
splitter ,AT- attenuator, P- polarizer, L4- $\protect\lambda$/4
plate, Rb - Rubidium cell in Helmholtz coils,Fl- fluorescence
detector, PD- photodiode, M- 100 \% mirror, Ref- reference Rb cell,
G- generator / booster,A$_{in}$ - analog in, A$_{out}$ - analog out.
} \label{f1_exp}
\end{figure}

\subsection{Experimental details}

\begin{figure}[tbp]
\includegraphics[width=5cm]{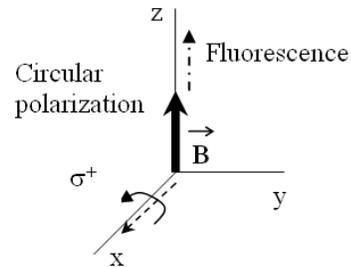}
\caption{Excitation light propagation, magnetic field and
fluorescence detection geometry.} \label{f2_light}
\end{figure}

A schematic drawing of the experimental setup is shown in Fig.
\ref{f1_exp}. A radiation beam from the solitary laser diode
(Sanyo$^{\mathrm{TM}}$ DL-7140-201 W)  was directed into the $1$
cm-long glass cell containing natural rubidium. The temperature of
the cell was $T_{cell}=24$ C ($N_{Rb}$ = $8.6\times 10^{9}$
cm$^{-3}$). The measured output power was $45$ mW at the Rb D$_{2}$
line, the spectral linewidth was $\sim 15$ MHz, and the  laser beam
diameter was $1.5$ mm. The cell was placed in $3$ mutually
orthogonal pairs of Helmholtz coils without metal shielding, which
reduced the \textit{dc} magnetic field to $\sim 10$ mG and could
apply a magnetic field of  up to $80$ G  in a chosen direction. The
resonant fluorescence emerging from the cell was detected by a
photodiode placed at an angle of $90$ degrees to the laser beam,
$14$ cm from the cell. The detection solid angle was $0.004$ srad.
The total intensity of the laser induced fluorescence was measured
without spectral or polarization selection. It was possible to
simultaneously record also the transmitted signal and the saturated
absorption signal (branched to an auxiliary setup, as shown in Fig.
\ref{f1_exp}). Control of the diode laser injection current and
hence, the radiation frequency, as well as data acquisition were
done by means of virtual function generator and a multi-channel
oscilloscope, with the help of a
National Instruments$^{\mathrm{%
TM}}$ DAQ board installed in the PC. The software was written in
LabVIEW$^{TM}$ . The measurements were performed by linearly
scanning the laser frequency over a $6$ GHz range around the Rb
D$_{2}$
line, covering the $^{87}$Rb $F_{g}=2\rightarrow F_{e}=1,2,3$, $^{85}$Rb $%
F_{g}=3\rightarrow F_{e}=2,3,4,$ and $^{85}$Rb $F_{g}=2\rightarrow
F_{e}=1,2,3$ Doppler-broadened and partially overlapping
transitions. The typical duration of a one-way scan was $5$ s.
Within this time period, the magnetic field is periodically scanned
by 120 triangular bipolar pulses, each with a duration of   of
$41.5$ ms. The number of measurement points per frequency scan was
$50000$ (over $400$ points per magnetic field scan). The fluorescence measurement in one scanning point takes $%
100$ $\mu $s. This measurement time interval is sufficiently large
to allow good signal resolution and the development of a steady
state interaction regime. The mutual orientation of the laser
radiation, its polarization, and the direction of the magnetic field
is shown in Fig.\ref{f2_light}.

\subsection{Experimental Results}

\begin{figure}[tbp]
\includegraphics[width=7.5cm]{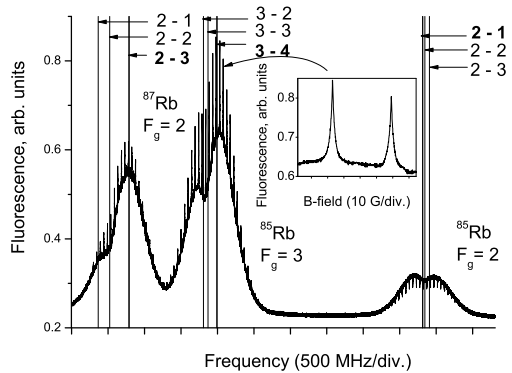}
\caption{Fluorescence signal for circular polarized excitation light
when simultaneously scanning the  mean frequency
$\bar{\protect\omega}$ of the laser light and the magnetic field B.}
\label{f3}
\end{figure}

Figure \ref{f3} show the \emph{double-scanning} fluorescence
excitation spectra for the circularly polarized excitation recorded
with laser intensity $I_{L}=1.25$
W/cm$^{2}$. We can clearly see fluorescence coming from the excitation of $%
^{87}$Rb atoms in the ground state level $F_{g}=2$, as well as
fluorescence
coming from the excitation of  $^{85}$Rb atoms in the ground state levels $F_{g}=3$ and $%
F_{g}=2$. Vertical lines in the graph show frequencies of atomic
transitions for atoms at rest. For example, line "$3-4$" shows the
position of the transition $F_{g}=3\rightarrow F_{e}=4$ for
$^{85}$Rb atoms. In this figure it can be seen that the structure of
the nonlinear Hanle signal depends on ground state HFS level from
which it was excited. The Hanle signals for excitation from the
$^{87}$Rb $F_{g}=2$ and from the $^{85}$Rb $F_{g}=3$ levels exhibit
sharp, high-contrast peaks in the vicinity of zero magnetic field.
The inset in Fig. \ref{f3} shows the dependance of the fluorescence
on the magnetic field  in the vicinity of these peaks. The width of
these high-contrast peaks is about $2$ G. In contrast, the
fluorescence excited from the ground state HFS\ level $F_{g}=2$ of
$^{85}$Rb exhibits dips in the vicinity of zero magnetic field of
approximately the same width as the previously discussed peaks. For
each of the three Doppler-overlapped groups of the three HFS
transitions, sub-Doppler dips appear at high laser intensity,
located at the frequency positions of the two crossover resonances
linked by the cycling
transitions. As was shown (but not comprehensively explained) in \cite%
{Pap99}, these dips arise even if precautions are taken to eliminate
the  backward-reflected beam from the cell, as was done also in this
work.

Similar spectra to the double-scanning fluorescence excitation
spectra depicted in Fig. \ref{f3} were taken in the laser intensity
range from $70$ mW/cm$^{2}$ up to $1.25$ W/cm$^{2}$.

\section{Theoretical Model}

\subsection{Outline of the model}

In order to build a model of the nonlinear Hanle effect in Rb atoms
in a cell, we will explore the concept of the density matrix of an
atomic ensemble. The diagonal elements of the density matrix $\rho
_{ii}$ of an atomic ensemble describe the population of a certain
atomic level $i$, and the non-diagonal elements $\rho _{ij}$
describe coherences created between the levels $i$ and $j$. In our
particular case the level in question are magnetic sublevels of a
certain HFS\ level. If atoms are excited from the ground state HFS
level $g$ to the excited state HFS level $e$, then the density
matrix consists of elements $\rho _{g_{i}g_{j}}$ and $\rho
_{e_{i}e_{j}}$, called Zeeman coherences, as well as
"cross-elements" $\rho _{g_{i}e_{j}}$, called optical coherences.

The optical Bloch equations (OBEs) can be written as \cite{Ste05}:
\begin{equation}
i\hbar \dfrac{\partial \rho }{\partial t}=\left[ \widehat{H},\rho \right]
+i\hbar \widehat{R}\rho ,  \label{1}
\end{equation}%
where the operator $\widehat{R}$ represents the relaxation matrix.
If an atom interacts with laser light and an external \emph{dc}
magnetic field, we can
write the Hamiltonian $\widehat{H}=\widehat{H}_{0}+\widehat{H}_{B}+\widehat{%
V}$. $\widehat{H}_{0}$ is the unperturbed atomic Hamiltonian, which
depends on the internal atomic coordinates, $\widehat{H}_{B}$ is the
Hamiltonian of the atomic interaction with the magnetic field, and $\widehat{V%
}=-\widehat{\mathbf{d}}\cdot \mathbf{E}\left( t\right) $, the dipole
interaction operator, where $\widehat{\mathbf{d}}$\ is the electric
dipole operator and $\mathbf{E}\left( t\right) $, the electric field
of the excitation light.

To use the OBEs to describe the interaction of alkali atoms with
laser radiation in the presence of a \emph{dc} magnetic field, we
describe the light
classically as a time dependent electric field of a definite polarization $%
\mathbf{e}$:%
\begin{equation}
\mathbf{E}\left( t\right) =\varepsilon \left( t\right) \mathbf{e}%
+\varepsilon ^{\ast }\left( t\right) \mathbf{e}^{\ast }  \label{2a}
\end{equation}%
\begin{equation}
\varepsilon (t)=\left\vert \varepsilon _{\overline{\omega }}\right\vert
e^{-i\Phi \left( t\right) -i\left( \overline{\omega }-\mathbf{k}_{\overline{%
\omega }}\mathbf{v}\right) t},  \label{2b}
\end{equation}%
where $\overline{\omega }$\ is the center frequency of the spectrum
and  $\Phi \left( t\right) $ is the fluctuating phase, which gives
the spectrum a finite bandwidth $\triangle \omega $. In this model
the line shape of the exciting light is Lorentzian with line-width
$\triangle \omega $. The atoms move with
a definite velocity $\mathbf{v}$, which causes the shift $\overline{\omega }-%
\mathbf{k}_{\overline{\omega }}\mathbf{v}$ in the laser frequency due to the
Doppler effect, where $\mathbf{k}_{\overline{\omega }}$ is the wave vector
of the excitation light.

The dipole matrix element that couples the $i$ sublevel with the $j$
sublevel can be written as: $d_{ij}=\langle i|\mathbf{d}\cdot
\mathbf{e}|j\rangle $. In the external magnetic field, sublevels are
mixed so that each sublevel $i$ with magnetic quantum number $M$ and
and other quantum numbers labeled as $\xi $ is mixture of different
hyperfine states $|F~M\rangle $ with mixing coefficients
$C_{i,F,M}$:
\begin{equation}
|i\rangle =|\xi ~M\rangle =\sum_{F}C_{i,F,M}|F~M\rangle   \label{3}
\end{equation}%
The mixing coefficients $C_{i,F,M}$ are obtained as the eigenvectors of the
Hamiltonian matrix of a fine structure state in the external magnetic field.

The dipole transition matrix elements $\langle
F_{k}M_{k}|\mathbf{d}\cdot \mathbf{e}|F_{l}M_{l}\rangle $ should be
expanded further using angular momentum algebra, including the
Wigner -- Eckart theorem and the fact that the dipole operator acts
only on the electronic part of the hyperfine state, which
consists of electronic and nuclear angular momentum (see, for example, \cite%
{Auz05}).

\subsection{Rate equations}

The rate equations for Zeeman coherences are developed by applying
the rotating wave approximation to the optical Bloch equations,
using an adiabatic elimination procedure for the optical
coherences\cite{Ste05}, and then accounting realistically for the
fluctuating laser radiation by taking
statistical averages over the fluctuating light field phase ( \emph{%
the decorrelation approximation}), and assuming a specific phase
fluctuation model -- random phase jumps or continuous random phase
diffusion. As a result we arrive at the rate equations for Zeeman
coherences for the ground and excited state sublevels of atoms
\cite{Blu04}. In applying this approach to a case in which atoms are
excited only in the finite region corresponding to the laser beam
diameter we have to take into account transit relaxation. Then we
obtain the following result:
\begin{widetext}
\begin{center}
\begin{eqnarray}
\dfrac{\partial \rho _{g_{i}g_{j}}}{\partial t} &&=-i \omega
_{g_{i}g_{j}}\rho _{g_{i}g_{j}}-\gamma \rho _{g_{i}g_{j}}+\underset{e_{i}e_{j}}{%
\sum }\Gamma _{g_{i}g_{j}}^{e_{i}e_{j}}\rho _{e_{i}e_{j}}+\lambda
\delta
\left( g_{i},g_{j}\right)  \notag \\
&&+\dfrac{\left\vert \varepsilon _{\overline{\omega }}\right\vert
^{2}}{\hbar ^{2}}\underset{e_{k},e_{m}}{\sum }\left(
\dfrac{1}{\Gamma _{R}+ i \Delta _{e_{m}g_{i}}}+\dfrac{1}{\Gamma
_{R}- i \Delta _{e_{k}g_{j}}}\right)
d_{g_{i}e_{k}}^{\ast }d_{e_{m}g_{j}}\rho _{e_{k}e_{m}}  \notag \\
&&-\dfrac{\left\vert \varepsilon _{\overline{\omega }}\right\vert ^{2}}{%
\hbar ^{2}}\underset{e_{k},g_{m}}{\sum }\left( \dfrac{1}{\Gamma
_{R}- i
\Delta _{e_{k}g_{j}}}d_{g_{i}e_{k}}^{\ast }d_{e_{k}g_{m}}\rho _{g_{m}g_{j}}+%
\dfrac{1}{\Gamma _{R}+ i \Delta _{e_{k}g_{i}}}d_{g_{m}e_{k}}^{\ast
}d_{e_{k}g_{j}}\rho _{g_{i}g_{m}}\right)  \label{7a}
\end{eqnarray}

\begin{eqnarray}
\dfrac{\partial \rho _{e_{i}e_{j}}}{\partial t} &&=- i \omega
_{e_{i}e_{j}}\rho _{e_{i}e_{j}}-\gamma \rho _{e_{i}e_{j}}-\Gamma
\rho
_{e_{i}e_{j}}  \notag \\
&&+\dfrac{\left\vert \varepsilon _{\overline{\omega }}\right\vert
^{2}}{\hbar ^{2}}\underset{g_{k},g_{m}}{\sum }\left( \dfrac{1}{
\Gamma _{R}- i \Delta _{e_{i}g_{m}} }+\dfrac{1}{ \Gamma _{R}+ i
\Delta _{e_{j}g_{k}} }\right) d_{e_{i}g_{k}}d_{g_{m}e_{j}}^{\ast
}\rho
_{g_{k}g_{m}}  \notag \\
&&-\dfrac{\left\vert \varepsilon _{\overline{\omega }}\right\vert ^{2}}{%
\hbar ^{2}}\underset{g_{k},e_{m}}{\sum }\left( \dfrac{1}{\Gamma
_{R}+ i
\Delta _{e_{j}g_{k}}}d_{e_{i}g_{k}}d_{g_{k}e_{m}}^{\ast }\rho _{e_{m}e_{j}}+%
\dfrac{1}{\Gamma _{R}- i \Delta _{e_{i}g_{k}}}%
d_{e_{m}g_{k}}d_{g_{k}e_{j}}^{\ast }\rho _{e_{i}e_{m}}\right),
\label{7b}
\end{eqnarray}
\end{center}
\end{widetext}
where $g_{i}$  denotes the ground state magnetic sublevel,  $%
e_{j}$ denotes the excited state sublevel, $\omega _{ij}=\left(
E_{i}-E_{j}\right)
/\hbar $, $d_{e_{i}g_{j}}=\langle e_{i}|\mathbf{d}\cdot \mathbf{e}%
|g_{j}\rangle $ is the transition dipole matrix element,
\begin{equation}
\Delta _{ij}=\bar{\omega}-\mathbf{k}_{\bar{\omega}}\mathbf{v}-\omega _{ij},
\end{equation}
$\Gamma _{R}=\frac{\Gamma }{2}+\frac{\Delta \omega }{2}+\gamma $,
$\Gamma $ is the relaxation rate of the excited level, $\gamma $ is
the  transit relaxation rate, and $\lambda $ is the transit
relaxation rate at which "fresh" atoms move into the interaction
region. The rate $\lambda $ at
which atoms are supplied into the interaction region can be estimated as $%
1/(2\pi \tau )$, where $\tau $ is time in which atom crosses the
laser beam due to thermal motion at a velocity $v$. It is assumed
that the atomic equilibrium density outside the interaction region
is normalized to $1$, which leads to $\lambda =\gamma $. $\Gamma
_{g_{i}g_{j}}^{e_{i}e_{j}}$ is the  rate at which excited state
coherences are transferred to the ground state as a result of
spontaneous transitions \cite{Auz05}. If the system is closed, all
excited state atoms in spontaneous transition return to the initial state, $%
\underset{g_{i}g_{j}}{\sum }\Gamma _{g_{i}g_{j}}^{e_{i}e_{j}}=\Gamma $.

We look at quasi stationary excitation conditions so that $\partial
\rho
_{g_{i}g_{j}}/\partial t=\partial \rho _{e_{i}e_{j}}/\partial t=0$. In (\ref%
{7a}) the first term on the right-hand-side of the equation
describes the destruction of the ground state Zeeman coherences due
to magnetic sublevel splitting in an external magnetic field. The
second term characterizes the destruction of ground state density
matrix due to transit relaxation. The next term shows the transfer
of population and coherences from the excited state to the ground
state due to spontaneous transitions. The fourth term describes how
the population of "fresh atoms"  is supplied to the initial state
from the regions outside of the laser beam in a process of transit
relaxation. The fifth term shows the influence of the induced
transitions to the ground state density matrix. And finally, the
last term describes the change of the ground state density matrix in
the process of light absorption.

The terms in the equations for the excited state density matrix, Eq.
(\ref{7b}), can be described in a similar way.  The first term shows
the destruction of Zeeman coherences by the external magnetic field.
The second term shows transit relaxation. The third term is
responsible for the spontaneous decay of the state. The fourth term
shows light absorption, and the last term describes the  influence
of the induced transitions on the density matrix of the excited
state.

For a multilevel system interacting with laser radiation, we can
define the
effective Rabi frequency in the form $\Omega =\dfrac{\left\vert \varepsilon _{%
\overline{\omega }}\right\vert }{\hbar }\left\langle
J_{e}\right\Vert d\left\Vert J_{g}\right\rangle $, where $J_{e}$ is
the angular momentum of the excited state fine structure level and
$J_{g}$ is the angular momentum of the ground state fine structure
level. The influence of the magnetic field appears directly in the
magnetic sublevel splitting $\omega _{ij}$ and indirectly in the
mixing coefficients $C_{i,F_{k},M_{i}}$ and $C_{j,F_{l},M_{j}}$ of
the dipole matrix elements $d_{ij}$.

By solving the rate equations as an algebraic system of linear
equations for $\rho _{g_{i}g_{j}}$ and $\rho _{e_{i}e_{j}}$ we get
the density matrix of the excited state. This matrix allows us to
obtain immediately the intensity of the fluorescence characterized
by the polarization vector $\mathbf{\tilde{e}} $ \cite{Auz05}:
\begin{equation}
I(\mathbf{\tilde{e}})=\tilde{I_{0}}%
\sum_{f_{i},e_{i},e_{j}}d_{f_{i}e_{j}}^{(ob)\ast
}d_{e_{i}f_{i}}^{(ob)}\rho _{e_{i}e_{j}},  \label{intensity}
\end{equation}

where $\tilde{I_{0}}$ is a proportionality coefficient. The dipole
transition matrix elements $d_{e_{i}f_{j}}^{(ob)}=\langle e_{i}|\mathbf{%
d\cdot \tilde{e}}|f_{j}\rangle $ characterize the dipole transition
from the excited state $e_{i}$ to some final state $f_{j}$ for the
transition on which the fluorescence is observed.

When we want to get the formula for the fluorescence which is
produced by an ensemble of atoms, we have to average the previously
written expression for the
fluorescence over the Doppler profile  $I(\mathbf{\tilde{e}})=I(\mathbf{\tilde{e}%
},\mathbf{k}_{\overline{\omega }}\mathbf{v},B)$, taking into account
 different velocity groups
$\mathbf{k}_{\overline{\omega }}\mathbf{v}$. If the total
fluorescence without discrimination of the polarization or frequency
is recorded, one needs to sum the fluorescence over all the
polarization components and all possible final state HFS levels.

\subsection{Theoretical results}

Our model was used to analyze experimental data obtained for atoms
excited with circularly polarized light at a geometry shown in Fig.
\ref{f2_light}.

In the experiment, atoms are excited at the resonance D$_{2}$ line
on the  fine structure transition $5S_{1/2}\leftarrow \rightarrow
5P_{3/2}$. The fine structure states are split and form the manifold
of hyperfine levels corresponding to nuclear spin $I=5/2$ and $3/2$
for $^{85}$Rb and $^{87}$Rb
respectively. The hyperfine constants for these levels can be found in \cite%
{Ari77}. In the signal simulation, the following numerical values of
the
parameters related to the experiment conditions were used: $\Gamma =6$ MHz, $%
\gamma =0.01$ MHz, $\Delta \omega =15$ MHz, $\Delta \nu _{D}=500$
MHz. We found that the model reproduces the measured signals
reasonably well for all frequencies of the excitation laser
including excitation rather far from the maximal absorption
frequency.

The total intensity of the fluorescence was simulated without
selecting the frequency or polarization, as was done in the
measurement. For the theoretical simulation we also assume that the
laser frequency does not change significantly during one magnetic
field scan.

In the calculation we found that the Rabi frequency $\Omega =250$ MHz gives
the best fit of experimental measurements obtained at the laser light
intensity $W=1250$ mW/cm$^{2}$.

\begin{center}
\begin{figure*}[tbp]
\includegraphics[width=5.8 cm]{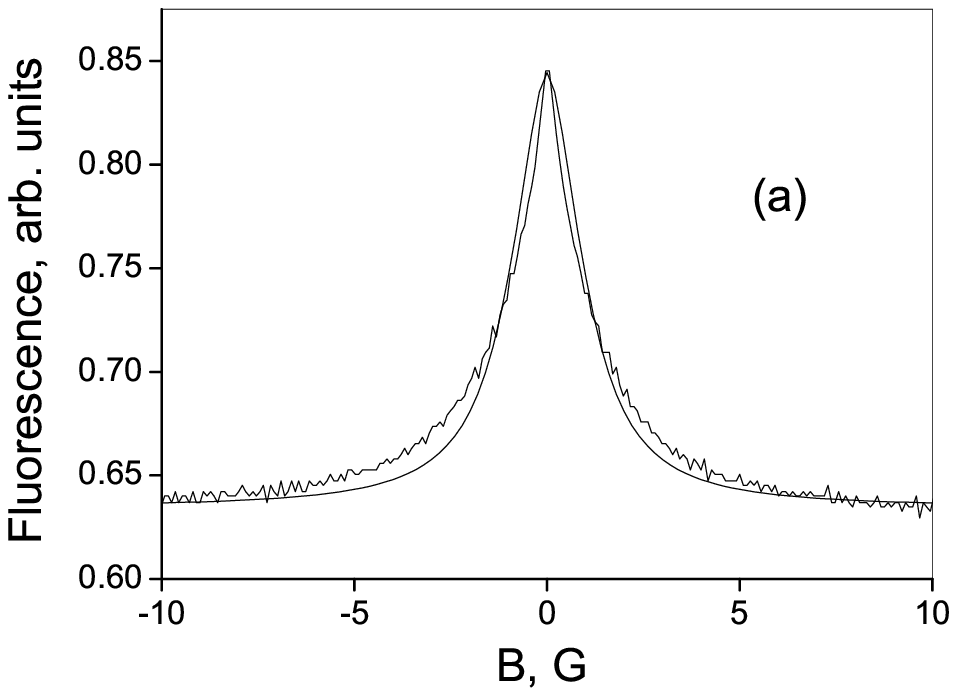} \includegraphics[width=5.8
cm]{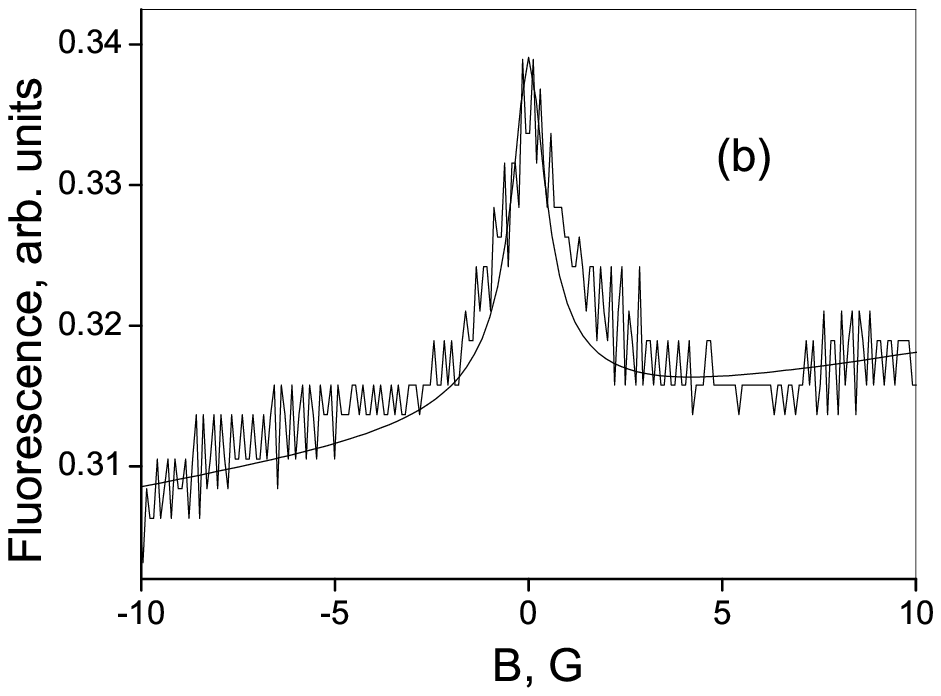} \includegraphics[width=5.8 cm]{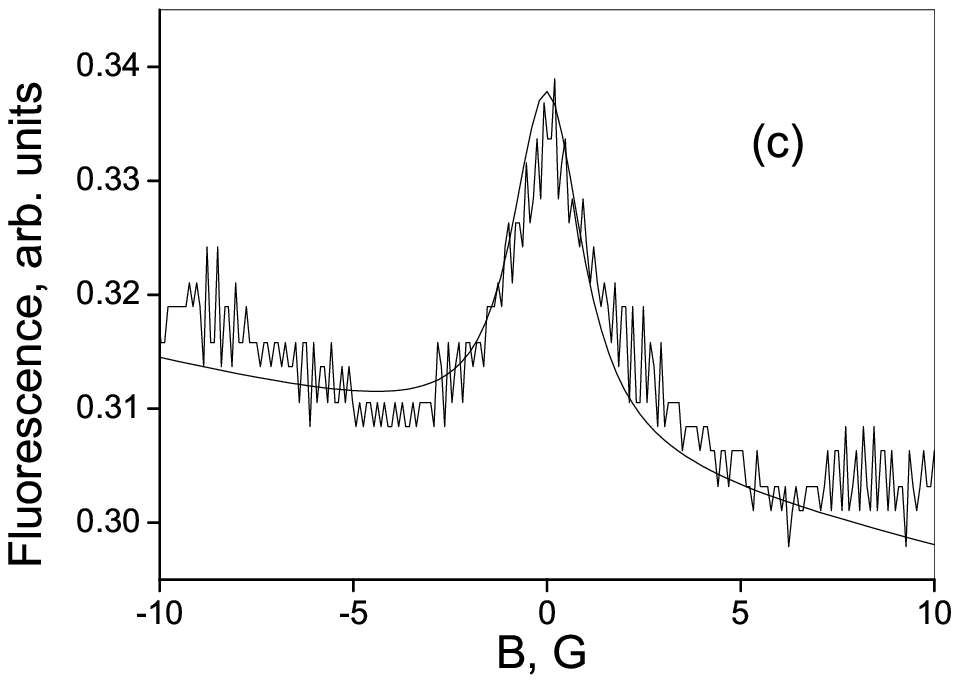}
\caption{Dependence of the fluorescence signal on the magnetic field
when the laser
light frequency $\bar{\protect\omega}$ is adjusted to the transition $\protect%
\omega _{34}$ ($a$), when $\bar{\protect\omega}=\protect\omega _{34}
- 625 MHz$ ($b$), and when $\bar{\protect\omega}=\protect\omega
_{34} + 390 MHz$ ($c$). The thin line represents the theoretical
simulation. } \label{f4}
\end{figure*}
\end{center}

The first set of data we analyzed was the dependance of the
fluorescence intensity  on the external magnetic field $B$ strength
for different excitation laser frequencies $\bar{\omega}$. In Fig.
\ref{f4} the dependance of the experimental and simulated
fluorescence signal on the magnetic field strength is plotted for
different laser frequencies in the region of excitation from the
F$_{g}$=3 hyperfine level of $^{85}$Rb (see Fig.\ref{f3}).
In case ($a$) it was assumed that the laser frequency $%
\bar{\omega}$ corresponds to the frequency $\omega _{34}$ of the HFS
cycling transition $F_{g}=3\rightarrow F_{e}=4$ of an atom at rest
when there is no magnetic field. In case ($b$) $\bar{\omega}=\omega
_{34}-625$ MHz, which corresponds to the laser frequency to the left
of the main peak of the
fluorescence excitation spectrum (see Fig. \ref{f3}). In case ($c$)  $%
\bar{\omega}=\omega _{34}+390$ MHz, which corresponds to the laser
frequency to the right of the main fluorescence peak.

\begin{center}
\begin{figure*}[tbp]
\includegraphics[width=5.8 cm]{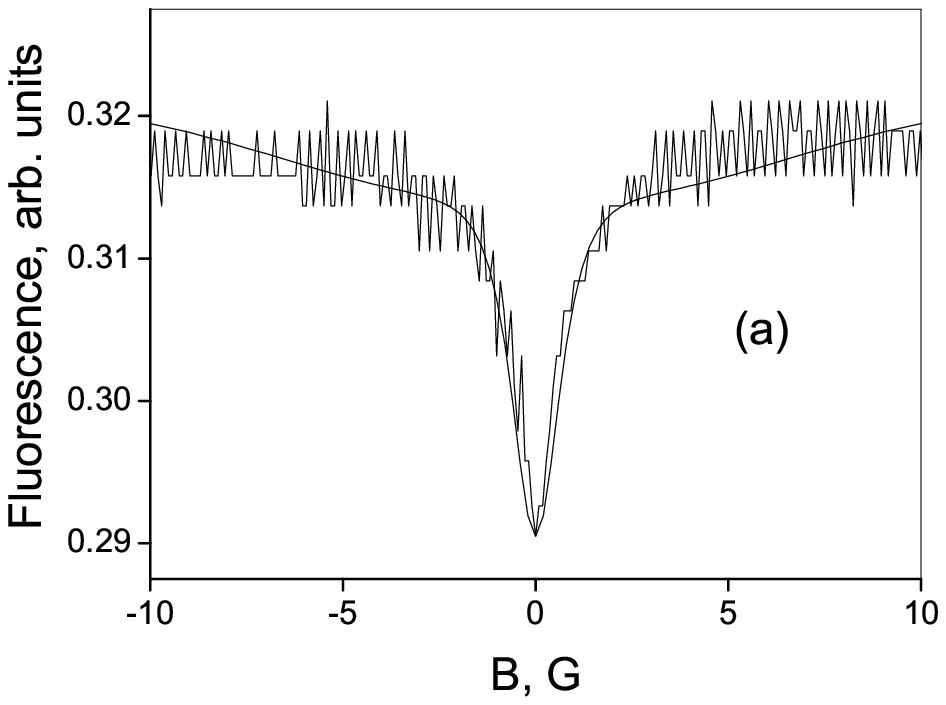} \includegraphics[width=5.8
cm]{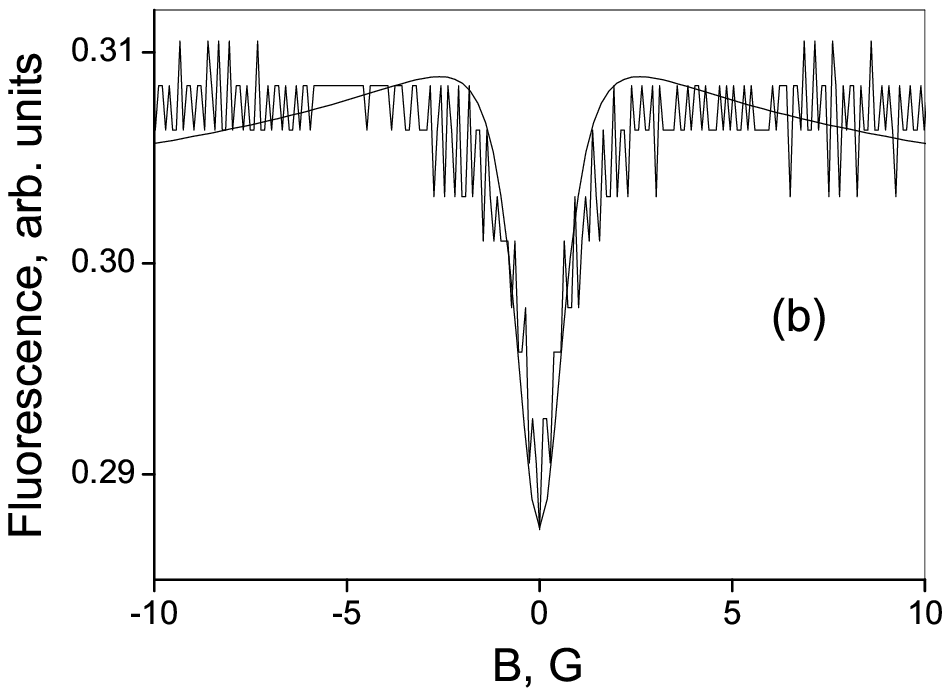} \includegraphics[width=5.8 cm]{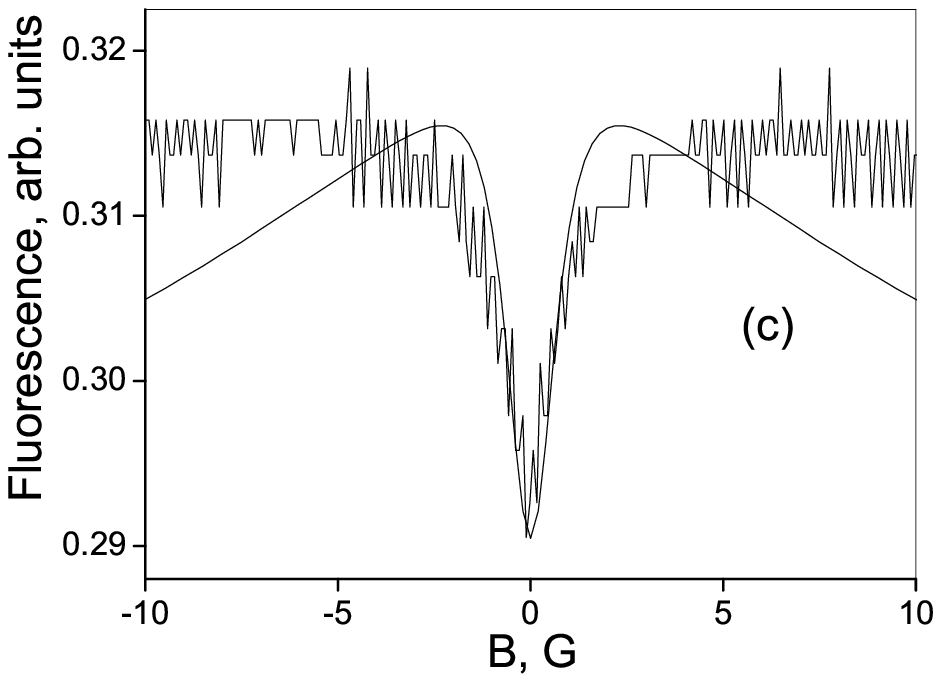}
\caption{Dependence of the fluorescence signal on the magnetic field
when laser
the frequency of the laser light $\bar{\protect\omega}$ is adjusted to the transition $\bar{%
\protect\omega}=\protect\omega _{34} + 2646 MHz$ ($a$), when $\bar{%
\protect\omega}=\protect\omega _{34} - 2793 MHz$ ($b$) and when $%
\bar{\protect\omega}=\protect\omega _{34} + 2935 MHz$ ($c$). The
thin line represents the theoretical simulation. } \label{f5}
\end{figure*}
\end{center}

The second set of the data we analyzed (see Fig. \ref{f5}) was
dependance of the  fluorescence on the external magnetic field $B$
when we are observing the excitation from the $^{85}$Rb ground state
hyperfine level $F_{g}=2$ (see Fig. \ref{f3}). In case ($a$) the
laser frequency was $\bar{\omega}=\omega _{34}+2646~$MHz, for
case($b$) it was $\bar{\omega}=\omega _{34}+2793$ MHz, and for case
($c$) $\bar{\omega}=\omega _{34}+2935$ MHz.

\begin{center}
\begin{figure*}[tbp]
\includegraphics[width=5.8 cm]{f4a.eps}
\includegraphics[width=5.8 cm]{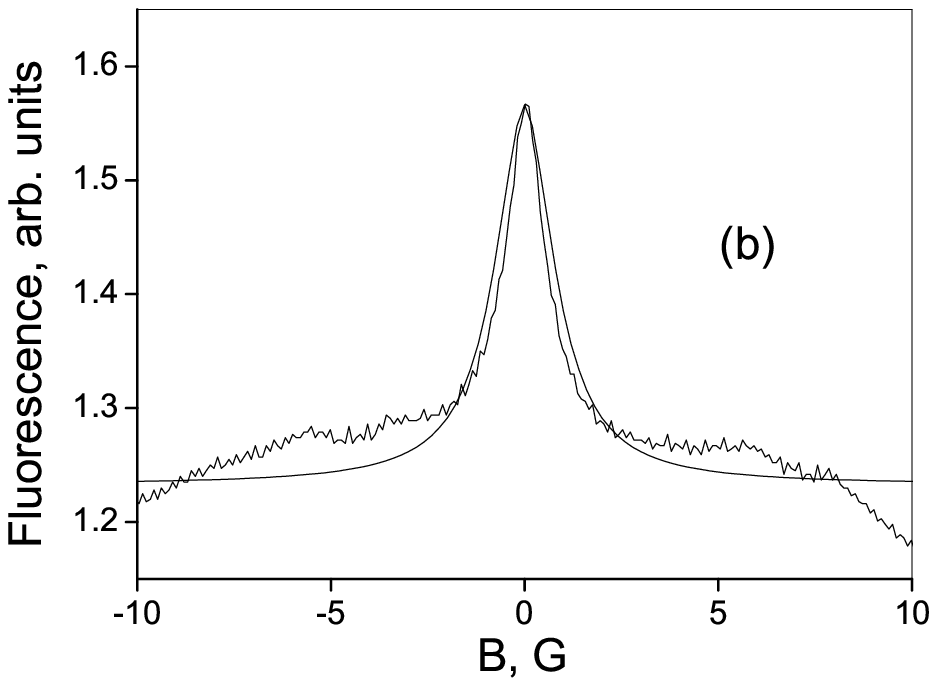}
\includegraphics[width=5.8 cm]{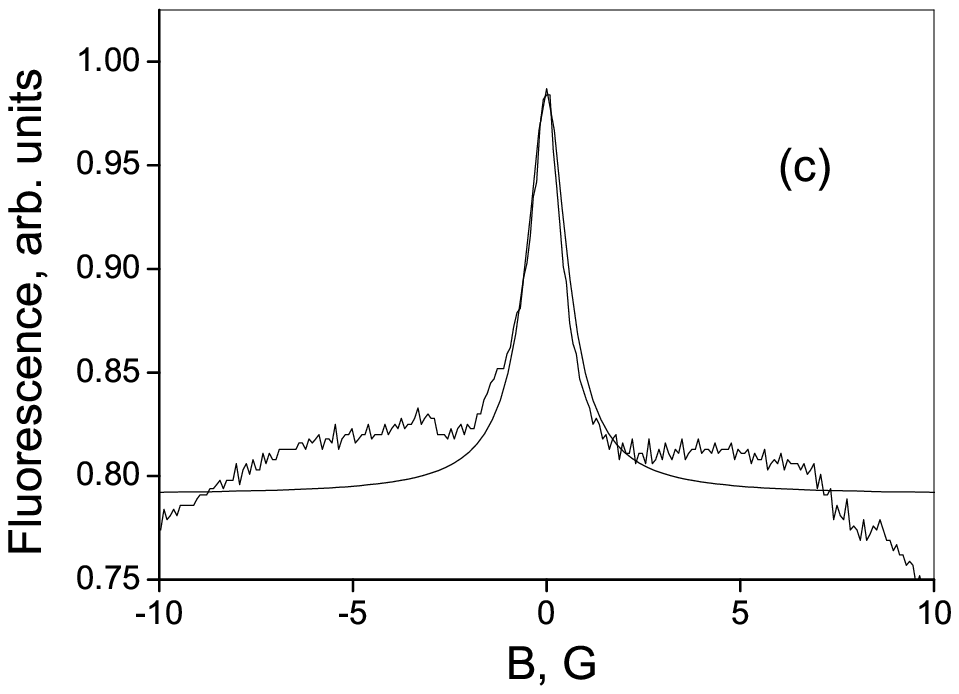}
\includegraphics[width=5.8 cm]{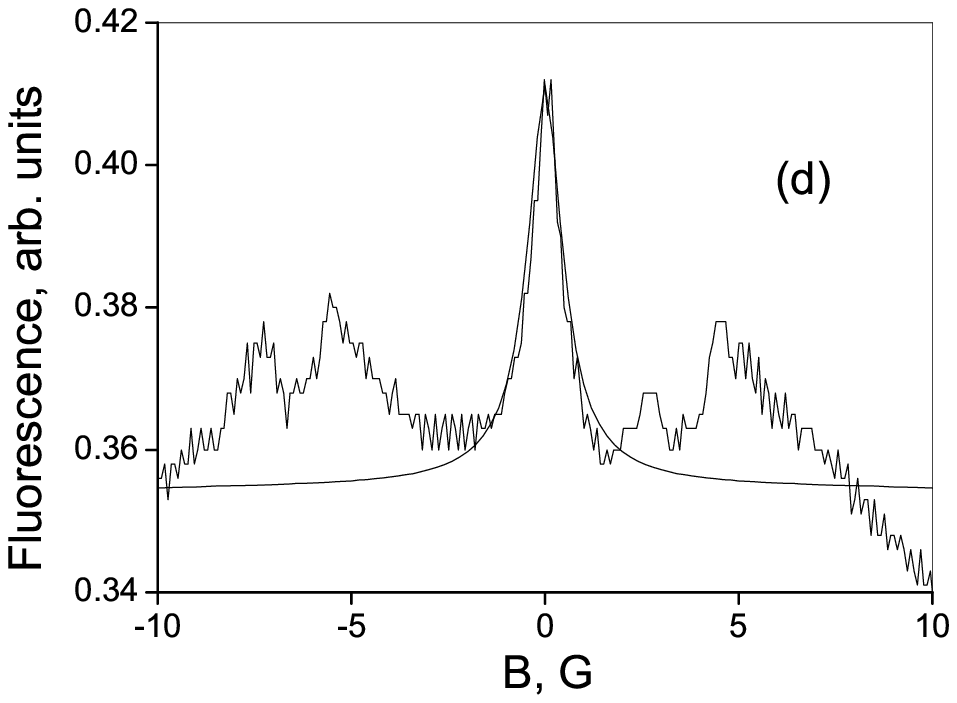}
\caption{Dependence of the fluorescence signal on the magnetic field
when the mean frequency of the laser light
$\bar{\protect\omega}=\protect\omega _{34}$ and the intensity $I$ of
the laser light is: ($a$) $I$ = 1250 mW/cm$^{2}$ ($\Omega$= 250
MHz); ($b$) $I$= 900 mW/cm$^{2}$ ($\Omega$= 215 MHz); ($c$) $I$= 330
mW/cm$^{2}$ ($\Omega$= 130 MHz); ($d$) $I$ = 70 mW/cm$^{2}$
($\Omega$= 60 MHz). The thin line represents the theoretical
simulation.} \label{f6}
\end{figure*}
\end{center}

The third set of data we analyzed was the dependance of the
fluorescence on the external magnetic field $B$ for different laser
light intensities. We chose to use a $B$-field scan that
corresponded to the laser frequency $\bar{\omega}=\omega _{34}$.
Since the intensity of the laser light should be proportional to
$\Omega ^{2}$ (see, for example, \cite{Auz04}), and since it was
found from our simulation that for $I=1250$ mW/cm$^{2}$, the
corresponding value of the Rabi frequency is $\Omega =250$ MHz, it
can be expected that for the laser intensities $I=900,330,$and $70$
mW/cm$^{2}$, which were also used in the experiment, the values of
the corresponding Rabi frequencies should be $\Omega =215,130,$ and
$60$ MHz, respectively. The results of the comparison of the
measured signals with the simulated curves are presented in Fig.
\ref{f6}.

\section{Analysis and discussion}

In our theoretical simulations as well as in the experimental
results, we see qualitatively different changes in the fluorescence
intensity for different excitation frequencies when the magnetic
field is scanned. For some excitation frequencies the peaks in the
vicinity of zero magnetic field are observed, while for other
excitation frequencies  dips are observed instead. In particular, in
the case when excitation occurs from the ground state HFS level
$F_{g}=3$ of $^{85}$Rb(see Fig. \ref{f4}), we see a peak at $B=0$.
This dip is called a
 \emph{bright resonance}. It typically occurs when the excitation
scheme $F_{g}\rightarrow F_{e}=F_{g}+1$ is executed. A detailed
explanation of the mechanisms that lead to the formation of bright
resonances is given in \cite{Aln01b,Ren01,Pap02}.

To use this mechanism to explain the peaks in the signals simulated
and measured for
excitation of $^{85}$Rb atoms from the $F_{g}=3$ and $^{87}$Rb atoms from the $%
F_{g}=2$ ground state HFS levels we have to demonstrate that the respective
transitions $F_{g}=3\rightarrow F_{e}=4$ and $F_{g}=2\rightarrow F_{e}=3$
give the main contribution to the fluorescence signal.

Let us perform this analysis for the absorption from the $F_{g}=3$
state to excited state levels $F_{e}=2,3,4$ of $^{85}$Rb. In the
present experimental conditions, we can spectrally resolve
ground-state HFS levels of Rb and cannot resolve excited-state HFS\
levels. Two points are important to understand the formation of
these bright resonances. First, for this excitation scheme the laser
line-width ($15$ MHz) is small in comparison to the mean energy
separation (around $100$ MHz) between excited state HFS levels. As a
result, we can assume that different atoms in different velocity
groups interact with laser light independently. Besides taking into
account the Doppler contour width (around $500$ MHz), we can assume
that for a rather large detuning of the laser frequency from the
exact resonance of any of the transitions $F_{g}=3\rightarrow
F_{e}=2,3,4$, there still will be atoms in some velocity group that
 will be in resonance with the laser excitation as a result of the Doppler shift.

Secondly, the ground state HFS levels for $^{85}$Rb have quantum
numbers $F_{g}=2,3$. The excited $P_{3/2}$ state has HFS levels with
quantum numbers $F_{e}=1,2,3,4$. For the transitions that start from
the spectrally resolved ground-state level $F_{g}=3$ only the
transition $F_{g}=3\rightarrow F_{e}=4$ is closed (cycling). This
means that atoms after the
absorption of a photon can return only to the initial ground-state level $%
F_{g}=3$ and cannot go to the other ground-state level $F_{g}=2$,
which does not interact with the laser light. The consequence of
this situation is that after several absorption and fluorescence
cycles, all atoms will be optically pumped to the $F_{g}=2$ level,
except atoms with the velocity group that corresponds to the
resonance with closed $F_{g}=3\rightarrow F_{e}=4$ transition. This
transition exhibits a bright resonance  in the magnetic field. This
is the reason why in the experiment, even when the laser radiation
is detuned from the exact $F_{g}=3\rightarrow F_{e}=4$ resonance to
the resonance with open transitions for the atom at rest, we still
observe bright resonances. This qualitative explanation of the
mechanism of resonance formation is supported by the solution of the
exact model.

In the case when excitation occurs for $^{85}$Rb from the ground-state HFS level $%
F_{g}=2$ (see Fig. \ref{f5}) we observe dips in the vicinity of
$B=0$. This dip is called a \emph{\ dark resonance}. It typically
occurs when the
excitation scheme is $F_{g}\rightarrow F_{e}=F_{g}-1$\cite%
{Aln01b,Ren01,Pap02}. For the excitation from the ground state HFS level $%
F_{g}=2$ in $^{85}$Rb the only closed transition is
$F_{g}=2\rightarrow F_{e}=1$. Similar to the previous case, one can
argue that  the dark resonances when atoms are excited from the
$F_{g}=2$ in $^{85}$Rb should be attributed to this transition at
any laser frequency within the Doppler profile.

\section{Concluding Remarks}

In summary, we can conclude that the double scanning technique is a
powerful tool to study simultaneously frequency and magnetic field
dependence of nonlinear zero field level-crossing signals, the
nonlinear Hanle effect. In this report we wanted to draw attention
to the extended possibilities offered by this technique. Together
with the improved model of the magneto-optical processes it allows
efficient measurements of these signals and provides a rather
detailed theoretical description of the signals that reproduces
experimentally measured signals with high accuracy.

\section{Acknowledgements}

This work was supported in part by grant $\# 0049$ of the Ministry
of Education and Science of Armenia, EU FP6 TOK Project LAMOL,
European Regional Development Fund project Nr. 2.5.1./000035/018 and
INTAS project Nr. 06 - 1000017 - 9001. The authors are grateful to
D. Sarkisyan for stimulating discussions. A. A. is grateful to the
European Social Fund for support.

\bibliographystyle{plain}

\begin{thebibliography}{19}

\bibitem{Mor91} G. Moruzzi and F. Strumia, \emph{Hanle Effect and Level - Crossing
Spectroscophy} (Plenum Press, New York Lon- don, 1991).
\bibitem{Ale93} E. Alexandrov, M. Chaika, and G. Khvostenko, \emph{Interference of
Atomic States} (Springer Verlag, New York, 1993).
\bibitem{Bud02} D. Budker, W.
Gawlik, D. F. Kimball, S. M. Rochester, V. V. Yashchuk, and A. Weis,
Rev. Mod. Phys. \textbf{74}, 1153 (2002).
\bibitem{Ale05} E. Alexandrov, M. Auzinsh, D.
Budker, D. Kimball, S. Rochester, and V. Yashchuk, Journal Of The
Optical Society Of America B - Optical Physics \textbf{22}, 7
(2005).
\bibitem{Pap02} A. Papoyan, M. Auzinsh, and K. Bergmann, Eur. Phys. J. D 21, 63 (2002).
\bibitem{Aln03} J. Alnis, K. Blushs, M. Auzinsh, S. Kennedy, N. Shafer- Ray, and
E. Abraham, Journal of Physics B-Atomic Molecular and Optical
Physics \textbf{36}, 1161 (2003).
\bibitem{And07} C. Andreeva, A. Atvars, M. Auzinsh, K. Bluss, S. Car-
leteva, L. Petrov, D. Sarkisya, and D. Slavov,  Eur. Phys. J. D. (to
be published)
\bibitem{Pap06} A.Papoyan and E. Gazazyan, Applied
Spectroscopy \textbf{60}, 1085 (2006).
\bibitem{Ste05} S. Stenholm, \emph{Foundations of Laser Spectroscopy} (Dover
Publications, Inc., Mineola, New York, 2005).
\bibitem{Blu04} K. Blush and M. Auzinsh, Phys.
Rev. A \textbf{69}, 063806 (2004).
\bibitem{Aln01} J. Alnis and M. Auzinsh, Phys. Rev. A
\textbf{63}, 023407 (2001).
\bibitem{Sar05} D. Sarkisyan, A. Papoyan, T. Varzhapetyan,
K. Blushs, and M. Auzinsh, J. Opt. Soc. Am. B \textbf{22}, 88
(2005).
\bibitem{Sar01} D. Sarkisyan, D. Bloch, A. Papoyan, and M. Ducloy, Optics
Communications \textbf{200}, 201 (2001).
\bibitem{Pap99} A. Papoyan, R. Unanyan, and K.
Bergmann, Verhand- lungen der Deutschen Physicalischen Gesellschaft
\textbf{44}, 63 (1999).
\bibitem{Auz05} M. Auzinsh and R. Ferber, \emph{Optical Polarization
of Molecules} (Cambridge University Press, Cambridge, 2005).
\bibitem{Ari77} E. Arimondo, M. Inguscio, and P. Violino, Rev. Mod. Phys. \textbf{49}, 31
(1977).
\bibitem{Auz04} M. Auzinsh, in \emph{Theory of chemical reaction dynamics},
edited by A. Lagana and G. Lendvay (Kluwer, New York, 2004), NATO
Science Series C, pp. 447 - 466.
\bibitem{Aln01b} J. Alnis and M. Auzinsh, Journal
of Physics B-Atomic Molecular and Optical Physics \textbf{34}, 3889
(2001).
\bibitem{Ren01} F. Renzoni, C. Zimmermann, P. Verkerk, and E. Ari- mondo,
Journal of Optics B-Quantum and Semiclassical Optics \textbf{3}, S7
(2001).

\end{thebibliography}

\end{document}